\documentclass[sigconf]{acmart}
\usepackage[utf8]{inputenc}
\usepackage{booktabs}
\usepackage{multirow}
\usepackage{ctable}
\usepackage{tabularx}
\usepackage{graphicx}
\usepackage{array}
\usepackage[all]{nowidow}
\usepackage{balance}
\usepackage{xspace}
\usepackage{url}
\usepackage{adjustbox}
\copyrightyear{2018}
\acmYear{}
\setcopyright{acmcopyright}
\acmConference[]{}{}{}
\acmBooktitle{}
\acmPrice{15.00}
\acmDOI{}
\acmISBN{}

\newcommand{\omt}[1]{}
\newcommand{\firstmention}[1]{{\bf #1}}
\newcommand{\myparagraph}[1]{\vspace{0.3\baselineskip}\noindent{\textbf{#1}}.~}

\newcommand{\topical}{Sub-Topic\xspace}
\newcommand{\nonRelevant}{Not-Relevant\xspace}
\newcommand{\docLength}{Doc-Length\xspace}
\newcommand{\incQueryTerms}{Query-Terms\xspace}
\newcommand{\herding}{Herding\xspace}
\newcommand{\biasing}{Biasing\xspace}
\newcommand{\lmart}{LambdaMART\xspace}
\newcommand{\RM}{RM1\xspace}
\newcommand{\control}{Control\xspace}

\newcommand{\stb}{STB\xspace}
\newcommand{\sth}{STH\xspace}
\newcommand{\nrh}{NRH\xspace}
\newcommand{\dlh}{DLH\xspace}
\newcommand{\qth}{QTH\xspace}

\newcommand{\query}{q}
\newcommand{\topic}{T}
\newcommand{\subTopicParm}[1]{\topic^{sub}_{#1}}
\newcommand{\subTopic}{\topic^{sub}}
\newcommand{\doc}{d}

\newcommand{\arbTerm}{w}

\newcommand{\simFn}{Score}

\newcommand{\arbSet}{\mathcal{S}_{\subTopicParm{i}}}

\newcommand{\set}[1]{\{#1\}}
\newcommand{\definedas}{\stackrel{def}{=}}

\newcommand{\crossEnt}{CE}
\newcommand{\ce}[2]{\crossEnt(#1 \; \vert\vert \,\, #2)}

\newcommand{\prob}{p}

\newcommand{\condArbP}[3]{\ensuremath{#1(#2 \vert #3)}}
\newcommand{\condP}[2]{\condArbP{\prob}{#1}{#2}}

\newcommand{\relModel}{R\xspace}

\newcommand{\figHeight}{3cm}
\newcommand{\figWidth}{6.3cm}

\newcommand{\queryCover}{QueryCover\xspace}
\newcommand{\fracQuery}{FracQuery\xspace}
\newcommand{\pass}{P\xspace}
\newcommand{\act}{A\xspace}
\newcommand{\stba}{\stb-\act}
\newcommand{\stha}{\sth-\act}
\newcommand{\stbp}{\stb-\pass}
\newcommand{\sthp}{\sth-\pass}

\copyrightyear{2021}
\acmYear{2021}
\setcopyright{acmlicensed}\acmConference[CIKM '21]{Proceedings of the 30th
ACM International Conference on Information and Knowledge
Management}{November 1--5, 2021}{Virtual Event, QLD, Australia}
\acmBooktitle{Proceedings of the 30th ACM International Conference on
Information and Knowledge Management (CIKM '21), November 1--5, 2021,
Virtual Event, QLD, Australia}
\acmPrice{15.00}
\acmDOI{10.1145/nnnnnnn.nnnnnnn}
\acmISBN{978-x-xxx-xxxx-x/YY/MM}

\begin{document}

\title{Driving the Herd: Search Engines as Content Influencers}

\author{Gregory Goren}
\authornote{Work done while at the Technion.}
  \email{ggoren@ebay.com}
\affiliation{%
  \institution{eBay Research}
}

\author{Oren Kurland}
\email{kurland@technion.ac.il}
\affiliation{%
  \institution{Technion}
}

\author{Moshe Tennenholtz}
\email{moshet@ie.technion.ac.il}
\affiliation{%
  \institution{Technion}
}

\author{Fiana Raiber}
\email{fiana@yahooinc.com}
\affiliation{%
  \institution{Yahoo Research}
}

\begin{abstract}
In competitive search settings such as the Web, many documents'
authors (publishers) opt to have their documents highly ranked for
some queries. To this end, they modify the documents --- specifically,
their content --- in response to induced rankings. Thus, the search
engine affects the content in the corpus via its ranking decisions.
We present a first study of the ability of search engines to drive
pre-defined, targeted, content effects in the corpus using simple
techniques. The first is based on the {\em herding} phenomenon --- a
celebrated result from the economics literature --- and the second is
based on biasing the relevance ranking function. The types of content
effects we study are either topical or touch on specific document
properties --- length and inclusion of query terms. Analysis of
ranking competitions we organized between incentivized publishers
shows that the types of content effects we target can indeed be
attained by applying our suggested techniques. These findings have
important implications with regard to the role of search engines in
shaping the corpus.
    \end{abstract}


\maketitle

\section{Introduction}
\label{sec:intro}

Search engines are mediators \cite{Tennenholtz+Kurland:19a}: they connect, using a query and a ranking function, users that have information needs with
content in the corpus. It is therefore not a surprise that search
engines have been traditionally perceived as ``passive observers'' of
the eco system they operate in. That is, they
search the corpus on behalf of users but do not actively affect the
corpus or document authors. This is indeed the
reality in search settings such as library archives and enterprise collections.


In large-scale adversarial search settings such as the Web, search
engines are far from being passive observers. To begin with, authors of Web pages --- henceforth referred to as publishers --- are
affected by induced rankings. Specifically, users pay most attention
to top-retrieved documents \cite{Joachims+al:05a} which has direct
effect on publishers' exposure
\cite{Yang+Stoyanovich:17a,Zehlike+al:17a,Castillo:18a,Biega+al:18a,Singh+al:18a}.


Rankings in adversarial (competitive) retrieval settings such
as the Web have additional effects on publishers. Specifically,
many publishers often change the content of their documents to have them
highly ranked in response to queries of interest --- a practice named search engine optimization (SEO). Thus, the ranking incentives of publishers, together with the search engine's ranking function, affect the content in the corpus. For example, it was shown, using game theoretical analysis, that applying common
relevance ranking functions\footnote{Specifically, 
  functions based on the
  probability ranking principle (PRP) \cite{Robertson:77a}: documents
  are ranked by their relevance probability. This holds for most relevance ranking functions.}
  in competitive retrieval settings results in decreased topical
coverage in the corpus \cite{Basat+al:17a}. One reason is
that a common
strategy of publishers is to mimic competing documents that are ranked higher \cite{Raifer+al:17a}, thereby potentially reducing topical diversity.

Despite their far reaching --- societal and other --- implications,
the effects of rankings induced by search engines on content in the
corpus have attracted very little research attention
\cite{Basat+al:17a,Raifer+al:17a}. Indeed, the large body of prior
work on studying content changes in the Web was performed regardless of
ranking effects (e.g., \cite{Radinsky+Bennett:13a,Radinski+al:13a,Acio+al:16a}).

We present the first
study, to the best of our knowledge, of the potential ability of
search engines to shape the content of documents in a corpus in {\em
  specific pre-defined} ways via the rankings they induce. To
demonstrate this ability, we explore a few types of content effects on
the corpus and techniques that a search engine can apply to drive
these effects. One of these techniques is essentially an example in the relevance ranking
domain of the celebrated {\em herding model} from the economics
literature \cite{Banerjee,Bikhchandani,SmithSorensen}.

Some of the content effects we study are
topical. That is, the search engine affects the coverage of topics in
documents, and to the extreme, the availability of content pertaining
to selected information needs. Other types of content effects touch on
specific document properties --- length and inclusion of query terms.

To empirically evaluate the content effects and the techniques for
driving them, and inspired by recent work on publishers' SEO
strategies \cite{Raifer+al:17a}, we organized content-ranking
competitions between students\footnote{The dataset is available at \url{https://github.com/herdingcikm/herding_data}.}. These competitions were approved
by international and institutional ethics committees. The students
produced and changed documents throughout time in response to induced
rankings so as to promote them in future rankings.  The empirical
findings that emerged from the competitions' analysis are quite
striking: search engines have an incredible power to drive pre-defined
targeted content effects in the corpus; specifically, using the
suggested techniques and with respect to the content-effect types we
studied.

It is well known that rankings
induced by search engines affect the corpus content since publishers are
often incentivized to have their documents highly ranked. However, this is a general observation with no concrete realization --- specifically, in terms of connecting the search engine ranking decisions with their corpus effects.
We present a novel
concrete realization: pre-defined, specific, content effects in the
corpus can be relatively easily driven by applying simple techniques.

It is also important to point out that {\em there is no reason for
  search engines to intentionally drive content effects.}  However,
their ability to do so due to the herding phenomenon has two direct
consequences. First, biases of the ranking function --- e.g., due to
biases in training data --- together with the herding phenomenon can
lead to unwarranted content effects on the corpus. A case in point,
using the cosine measure to rank documents in the vector space model
is known to lead to a bias in favor of short documents
\cite{Singhal+Buckley+Mitra:96a}. Now, we show in our experiments that
positioning very short documents at the highest ranks of the retrieved
document lists leads, due to the herding phenomenon, to a document
length decrease effect in a competitive retrieval setting.  Second, the
ability of a search engine to drive specific content effects can
potentially be abused by publishers who are interested in driving such
effects. Recent progress in language generation methods
facilitates opportunities for such an abuse. We discuss this issue in Section \ref{sec:discuss}. Hence, whenever we
write that the search engine can apply the techniques to drive
content effects we mean that it has the {\em ability} to do so. The
operational activation of this ability is either due to inherent biases of ranking functions or abuse by publishers.



In summary, we demonstrate the relative ease by which search engines
in competitive search settings can drive pre-defined, targeted,
content effects in the corpus due to ranking incentives of
publishers. This {concrete} realization of the effects of
ranking decisions on corpus content has important implications which
we discuss. The ranking-competitions dataset we created can facilitate further research on ranking effects in
competitive settings.

\section{Related Work}
\label{sec:rel}
Work on adversarial retrieval has mainly focused on different types of
spamming and methods to address them
\cite{AIRWeb,Gyongyi+Molina:05a,Castillo+Davison:10a}. In contrast, we
study the ability of search engines to affect corpus content.


Deception and spread of misinformation in content-based platforms,
specifically in social networks, was the subject of many studies
(e.g., \cite{Guillory+Hancock:12a,Bliss+al:20a}). Our focus is different: the connection between search engines' ranking decisions and content effects on the corpus.

The Facebook experiment
\cite{Kramer2014ExperimentalEO} showed that the sentiment of content
promoted in users' feeds affected the sentiment of posts the users
wrote. The search setting we address is different,
and our focus is on other types of content effects.

There are many studies of the interactions between search engines and their users \cite{White:16a}, including those of the effects of the engines on
users' behavior via the advertisements they promote
\cite{yomtov+al:16a,yomtov+al:18a}. In contrast, we study the effects
of search engines on the corpus via the rankings they induce.


There is work on ranking fairness with respect to publishers'
representation in top-retrieved results (e.g.,
\cite{Yang+Stoyanovich:17a,Zehlike+al:17a,Biega+al:18a,Singh+al:18a}). Our focus is on the effects of rankings on documents in the
corpus.


Herding of publishers, which is an economic/game-theoretic phenomenon,  
can already be observed when publishers compete on their relative ranking
with respect to a single query \cite{Raifer+al:17a}. This phenomenon occurs due to
uncertainty about the ranking function. Some recent work on search engines and recommendation systems deals with other game theoretic effects, resembling
the ones that appear in facility location games \cite{Basat+al:17a,Ben-PoratT18}. As in the single-query setting that we address in this paper, it
was shown that due to competition, publishers will be inclined to write
on similar topics \cite{Basat+al:17a,Ben-PoratT18}. This finding is reminiscent of the phenomenon that facilities of
competitors are located in similar locations in equilibrium. Interestingly, the finding holds for a set of queries even if publishers have 
full information of the ranking/recommendation function.

Raifer et al. \cite{Raifer+al:17a} showed that publishers in ranking
competitions tend to mimic content in documents highly ranked in the
past for the same query. This is evidence for the
general herding effect. We study the herding effect from the
perspective of leading to specific content effects and the ability to
actively drive them.



\setlength{\abovedisplayskip}{2pt}
\setlength{\belowdisplayskip}{2pt}
\setlength{\abovedisplayshortskip}{2pt}
\setlength{\belowdisplayshortskip}{2pt} 
\allowdisplaybreaks

\section{Content Effects}
Our goal is to study the ability of a search engine to drive
pre-defined types of content effects in a corpus upon which search is
performed. This ability relies on the fundamental characteristics of
any competitive search setting (e.g., the Web): some publishers (authors) of documents are incentivized to have their documents highly
ranked for queries of interest. Hence,
they respond to rankings induced for the queries by modifying their
documents --- a practice often referred to
as
search engine optimization (SEO) \cite{Gyongyi+Molina:05a}.
We focus on ``white-hat SEO'' {\em content}
modifications \cite{Gyongyi+Molina:05a}; i.e., legitimate
modifications of document content that are {\em not} considered
spamming, and more generally, that do not degrade document
quality. Yet, such modifications can certainly have negative effects on the search eco system as we discuss below.

Our treatment of content effects is on a per-query basis. That is, we
study how the search engine can affect, via the rankings it induces for a
given query, the content of documents whose publishers opt to promote
for this query. Obviously, publishers can strive to promote
their documents, simultaneously, for several queries. We leave the study of driving content effects via induced rankings for a set of queries for 
future work. 

In Section \ref{sec:types} we discuss the types of content effects we
explore. These are examples and do not constitute a
complete set of all potential effects. Section \ref{sec:means}
presents techniques for driving these effects. We use $\query$
and $\doc$
to denote a
query and a document, respectively.

\subsection{Types of Content Effect}
\label{sec:types}
The first type of content effect is with respect to topics discussed in documents. Suppose that query $\query$ represents the
topic (information need) $\topic$. We set as a goal for the search engine
to affect the treatment of $\topic$ in documents whose authors are
interested in rank promotion for $\query$. A concrete example we study here is
trying to bias the content in these documents towards a specific aspect, or
sub-topic, $\subTopic$ of $\topic$. Accordingly, we term this type of content effect \firstmention{\topical}.

For example, consider TREC topic 
\#167
from the ClueWeb09 
collection. The description of the information need (topic), $\topic$, is:
{\em ``Find information on Barbados history''}.
Two of the sub-topics, $\subTopicParm{1}$ and $\subTopicParm{2}$, for this topic are
described as: 
{\em ``What does the Barbados flag look like?''}
and 
{\em ```Suggest tourist activities in Barbados''},
respectively. The topic
title,
{``barbados''},
serves for the query $\query$. The question we explore below is how the
search engine can drive publishers interested in rank
promotion for $\query$ to focus on one of the two sub-topics.

While documents written on $\subTopicParm{1}$ and $\subTopicParm{2}$
are relevant to $\topic$, the sub-topic focus just described results in potential loss of
valuable information in the corpus. Namely, if one assumes that, a-priori, both
sub-topics are discussed in documents in the corpus, then the topical coverage in the corpus is reduced. The direct consequence is hurting, in
the long run, the effectiveness of any search intended for finding
information about the sub-topic which the publishers drift away from; i.e., the one not driven by the search engine.

The \topical effect results in different coverage of sub-topics of
$\topic$, but does not necessarily reduce the overall amount of
information in the corpus that pertains to $\topic$. We thereby take a
step forward and study the ability of the search engine to inflict more
harm in terms of topical coverage in the corpus; specifically, to
drive the reduction of the amount of content relevant to
$\topic$.\footnote{We write ``reduction'' and not ``elimination'' as
  we operate within the scope of a single query and other queries
  might target the same topic. Furthermore, there could be publishers
  with no rank incentives who are not affected by induced rankings.}
To this end, we introduce the \firstmention{\nonRelevant} content-type
effect: driving publishers with rank-promotion incentive for $\query$
to write documents not relevant to $\topic$. These documents can
naturally still include $\query$'s terms and exhibit other properties that result in having them highly ranked for $\query$, but their content cannot satisfy the information need about $\topic$.

The types of content effects we consider next touch on two 
fundamental building blocks of classical retrieval methods which rank
documents by surface-level similarity to the query
\cite{Fang+Zhai:05a}; namely, term frequency and document length. The \firstmention{\docLength} effect simply
refers to the impact on document length. The \firstmention{\incQueryTerms}
effect refers to the extent of occurrences of terms from the query,
$\query$, in a document whose publisher wants to rank-promote for
$\query$. Our goal here is to study whether the search engine can
drive publishers to reduce the number of query-term occurrences in
documents. This type of a change stands in clear contrast to the
common wisdom of publishers about relevance ranking functions; i.e.,
that they reward documents containing many occurrences of query
terms. Indeed, a common SEO technique is {\em keyword stuffing}:
adding query terms to documents \cite{Gyongyi+Molina:05a}. Thus, the
goal is to study whether a search engine can actually drive a content effect which contradicts the common belief of publishers about relevance ranking functions.

\subsection{Approaches for Driving Content Effects}
\label{sec:means}
We next describe approaches that a search
engine can employ to drive content effects; specifically, the
four types discussed above.

\subsubsection{Herding}
\label{sec:herding}
The first approach we consider is inspired by recent work on analyzing
the strategies employed by publishers who have rank-promotion
incentives \cite{Raifer+al:17a}. An important result of a
game theoretical analysis of the ``ranking competition'' between
publishers was as follows: a publisher who opts to promote her
document in rankings induced for query $\query$ should mimic documents that were highly ranked for $\query$ in the past
\cite{Raifer+al:17a}. Empirical analysis of ranking
competitions provided support for this
finding \cite{Raifer+al:17a}. The simple intuitive rationale behind this
``mimicking''  strategy is that given that the ranking function is not
known to publishers, the rankings they observe --- of their documents and others --- are essentially the
only (implicit) signal about the ranking function.

Interestingly, the mimicking strategy just described is a specific
example of the celebrated {\em herding model} from an exciting branch
of the economics literature
\cite{Banerjee,Bikhchandani,SmithSorensen}. The literature refers to a
paradigm known as {\em the wisdom of the crowd}; it suggests that a
population can aggregate knowledge from individuals to collectively
learn new things. A well established negative result from that
literature is that whenever agents have full observability of selected
actions of others, a herd may form on an inferior alternative. This
phenomenon is often referred to as an {\em information cascade}: an
information cascade occurs when an initial set of agents is
ill-informed. As a result, these agents take some inferior
action. Subsequent agents are then convinced that the aforementioned
action is optimal and so dismiss their own private information and
follow the herd by taking the inferior action as well. In our setting,
publishers are the agents who are ill-informed about the ranking
function and who follow the herd by mimicking documents highly ranked. The inferior alternatives they herd on can be, for example,
focusing on specific sub-topics or not producing relevant information
as in the content-type effects described in
Section \ref{sec:types}.

Following the observations about the mimicking strategy
\cite{Raifer+al:17a} and the herding phenomenon
\cite{Banerjee,Bikhchandani,SmithSorensen}, we apply the following
simple approach, denoted \firstmention{\herding}, for driving content effects. For query $\query$, we manually create a
document $\doc$ that manifests the type of content effect the search
engine opts to drive. Then, $\doc$ is positioned at the highest rank
of any ranking induced for $\query$, regardless of its actual
retrieval score. That is, the approach is agnostic to the ranking
function employed by the search engine. We note that using several manually created
documents per query and determining their positions is a study left for future
work.

For the \topical effect, the document $\doc$ is written so that it is
relevant to the sub-topic $\subTopic$ the search engine wants the
publishers to focus on; hence, the document is also relevant to the
topic $\topic$ that $\query$ represents . We take care that $\doc$ is not relevant to other (given)
sub-topics of $\topic$ so as to further highlight the potential topic
drift that the search engine drives.

For the \nonRelevant effect, we create a document which contains
$\query$'s terms, but is not relevant to the topic $\topic$ that
$\query$ represents. Such non-relevant documents are often highly
ranked by retrieval methods based on document-query
surface-level similarities.

For \docLength we create a short relevant
document, and for \incQueryTerms we create a relevant document which
does not contain any of the query terms. (This is mainly done by using
abbreviations and language that avoids using the query terms.)

It is important to make the following observations about the \herding
approach. First, if $\doc$ is kept as is at the first rank along
  time, then publishers will figure out that straightforward
mimicking of $\doc$ does not necessarily lead to improved ranking, and
more generally, that external intervention takes place.  This
is because $\doc$ is not necessarily assigned a (very) high retrieval
score, which will also be the case for documents mimicking it. To
address this issue, one should engineer $\doc$ --- using the full knowledge of the ranking function --- so that it not only
manifests the desired content effect, but it is also assigned a very
high retrieval score. Furthermore, $\doc$ can be changed along time,
and so does the rank at which it is positioned, to avoid suspicions of
the publishers. We leave the exploration of these directions for
future work and focus on the fundamental \herding principle. To that end, in the experiments with iterative
ranking competitions reported in Section \ref{sec:expResults}, we demonstrate
the impact of the approach when used for very few iterations (i.e., short time
span). This short span effect is  along the lines of the herding literature mentioned above
\cite{Banerjee,Bikhchandani,SmithSorensen} which deals with a single
decision made by agents, rather than a repeated one.

\omt{
\myparagraph{The publisher perspective} The \herding approach is
highly important also because it is a tool that can potentially be
abused by publishers to affect content in the corpus. That is, if a
publisher is interested in promoting content effects in the corpus
(i.e., to documents of other publishers), she can write a document
that manifests this effect and try to optimize it with respect to the
ranking function; i.e., this is a bi-modal optimization goal for
creating the document. If successful in having her document ranked first, the publisher can
potentially affect the content other publishers produce due to the
\herding effect. This observation has far reaching implications: while
there is a societal worry about the effect of search engines on their
users (e.g., due to having fake news being ranked high), the search
engine can potentially serve as a platform for publishers to affect
other publishers in adversarial ways --- e.g., spreading the fake news
among publishers due to the \herding effect.
}

\subsubsection{Biasing the Ranking Function}
\label{sec:funcBias}

In the \herding approach, publishers have an explicit example --- the
highest ranked document $\doc$ --- to follow towards
the desired content effect. The next approach, termed
\firstmention{\biasing}, is based on providing the publishers with an
implicit signal about the desired effect by biasing the relevance ranking
function. For example, documents can be rewarded also based on the extent
to which they manifest the desired content effect. Hence, changes that publishers introduce to documents and which are aligned with this effect will result in improved ranking.

A basic approach to biasing a ranking function is biasing the training
data used to learn the function. It is easy to bias
the training data for the four content-effect types discussed
in Section \ref{sec:types}: one can reward, via boosting of
relevance grades, the documents that manifest the effect. For the \topical effect, the relevance grades of documents relevant to the sub-topic the engine wants to focus on should be increased. For the \nonRelevant effect, non-relevant documents (but with high surface-level query similarity) should be assigned high relevance grades and relevant documents should be assigned low relevance grades. For the \docLength and \incQueryTerms effects, the relevance grades of short relevant documents and those of relevant documents which include very few occurrences of the query terms, respectively, should be increased.

As a proof
of concept, we present an alternative unsupervised approach to biasing the ranking function for the \topical effect. Specifically, since our focus is
on content effects, we devise a ranking function that is solely based on content. The retrieval scores assigned to documents by this function can be incorporated in feature-based learning-to-rank approaches so as to bias them \cite{Liu:11a}.

Let $\query$ be a query representing topic $\topic$, and suppose that the goal is to have publishers focus
on sub-topic $\subTopicParm{i}$ of $\topic$. We assume a set of documents
$\arbSet$ which were judged as relevant to $\subTopicParm{i}$.\footnote{This assumption corresponds to the scenario that the search engine received in the past the query $\query$ and relevance judgments were accordingly produced.} Then, we construct a unigram relevance language model from $\arbSet$ \cite{Lavrenko+Croft:03a}:

\begin{equation}
  \label{eq:relModel}
\condP{\arbTerm}{\relModel} \definedas \frac{1}{\vert \arbSet \vert}
\sum_{\doc' \in \arbSet} \condP{\arbTerm}{\doc'},
\end{equation}
where
$\condP{\arbTerm}{\relModel}$ is the probability assigned to term
$\arbTerm$ by the relevance model $\relModel$ and
$\condP{\arbTerm}{\doc'}$ is the probability assigned to $\arbTerm$ by
a (Dirichlet) smoothed language model induced from
$\doc'$.\footnote{Uniform weighting of documents for constructing a relevance model from true relevant documents is superior to other weighting approaches \cite{Lavrenko+Croft:03a,Raiber+Kurland:10a}. We use relevance model \#1 (RM1)
  and do not interpolate it with the original query model (RM3)
  as our experiments showed that the
  resultant effect is stronger for RM1.} To rank document $\doc$ using $\relModel$, we use the negative cross entropy:
\begin{equation}
  \label{eq:sim}
  \simFn(\doc;\query) \definedas
-\ce{\condP{\cdot}{\relModel}}{\condP{\cdot}{\doc}} = \sum_{\arbTerm}
\condP{\arbTerm}{\relModel} \log \condP{\arbTerm}{\doc}.
\end{equation}
Increased values of the cross entropy correspond to decreased language-model similarity.
Hence, a
document $\doc$ whose induced language model is similar to the
relevance model induced for sub-topic $\subTopicParm{i}$ will be
rewarded. This will presumably incentivize publishers to emphasize $\subTopicParm{i}$ in documents they want to promote for query $\query$.

\begin{table}[t]
\tabcolsep=0.11cm
\caption{\label{tab:groups} Summary of the ranking competitions.}
\hspace*{-.1in}
\scriptsize
\begin{tabular}{@{}lclllc@{}}
\toprule
& Competition & Effect & Approach & Ranking Function & \# Competitions\\
\midrule
\multirow{3}{*}{\begin{sideways}Part I\end{sideways}} 
& \control & None & None  & \lmart & $30$ \\
& \sth & \topical & \herding  & \lmart & $60$ \\
& \stb & \topical & \biasing  & \RM & $60$ \\
\midrule
\multirow{3}{*}{\begin{sideways}Part II\end{sideways}}
& \nrh & \nonRelevant & \herding  & \lmart & $30$ \\
& \dlh & \docLength & \herding  & \lmart & $30$ \\
& \qth & \incQueryTerms & \herding  & \lmart & $30$ \\
\bottomrule
\end{tabular}

\end{table}

We point out that the specific choice of a relevance model to
represent sub-topic $\subTopicParm{i}$ entails an implicit herding/mimicking
effect. That is, changing document $\doc$ to increase its
retrieval score means that its induced language model becomes more
similar to $\relModel$. Now, $\relModel$, as defined in Equation \ref{eq:relModel}, is an arithmetic centroid in the
simplex of the language models induced from the documents in $\arbSet$ which
represent $\subTopicParm{i}$. Hence, to promote their documents in a ranking induced for $\query$, publishers should essentially make them become more
similar to a pseudo document that represents $\subTopicParm{i}$. An additional perspective can be gained by
plugging Equation \ref{eq:relModel} in Equation \ref{eq:sim} and re-arranging the summations:
\begin{equation}
  \label{eq:docSim}
  \simFn(\doc;\query) \definedas - \frac{1}{\vert \arbSet \vert} \sum_{\doc' \in \arbSet}  \ce{\condP{\arbTerm}{\doc'}}{\condP{\arbTerm}{\doc}}.
  \end{equation}
That is, the retrieval score of $\doc$ is the average negative cross entropy
between language models induced from documents $\doc'$ in $\arbSet$ and $\doc$'s induced language model. Thus, promoting $\doc$ in a ranking means making
it more similar to these representatives of $\subTopicParm{i}$ which
the publisher is not aware of. This is in contrast to the explicit \herding effect from Section \ref{sec:herding} where publishers observe the document most highly ranked which was ``planted'' there by the search engine.

\begin{table*}[t]
\tabcolsep=0.11cm
\caption{\label{tab:docs} Example of documents created for topic \#167, which is represented by the query ``barbados''. $\topic$: `` Find information on Barbados history.'' $\subTopicParm{1}$: ``What does the Barbados flag look like?''. $\subTopicParm{2}$: ``Suggest tourist activities in Barbados.''}
\scriptsize
\begin{tabular}{@{}lp{15.5cm}@{}}
\toprule
Effect & Document \\
\midrule
Initial Document & \textit{``The island of Barbados is located at 13.4N and 54.4W and is situated in the western area of the North Atlantic Ocean and 100 kilometers east of the Windward Islands and the Caribbean Sea. The island is seen by most scientists as geologically unique as it was formed as a result of an amalgamation of two land masses over a period of many years. The peaceful Arawaks and the more ferocious Caribs were the first inhabitants of Barbados...''}  \\ \midrule
\topical ($\subTopicParm{1}$)  & \textit{``Barbados flag consists of a triband of two bands of ultramarine, which are said to stand for the ocean surrounding the country and the sky, separated by a golden middle band, which represents the sand. A black trident head, commonly called the broken trident, is centered in the golden band, and the fact that the staff is missing is significant. The trident symbol was taken from Barbados colonial badge, where the trident of Poseidon is shown with Britannia holding it...'' }\\ \midrule
\topical ($\subTopicParm{2}$)&  \textit{``Barbados is one of the most popular destinations for vacation in the Caribbean, due to its beautiful scenery and high standard of living. There are many excursions for travelers to this island nation to take advantage of, no matter what their travel interests may be. One place tourists will want to go in Barbados is Harrisons Cave...'' }\\
\midrule
\nonRelevant & \textit{``Barbados is known for two pirates of the Caribbean - Sam Lord and Stede Bonnet. Stede Bonnet - Known as the pirate gentleman, Stede Bonnet became one of the pirates of the Caribbean in a most unusual way! A retired British army major and well off plantation owner in Barbados, the middle aged Major Stede Bonnet suddenly turned to piracy in early 1717 and actually purchased his own pirate ship, an unheard of act among the pirates of the Caribbean!'' }\\ \midrule
\docLength & \textit{``The limestone rock has created the island of Barbados, and the land area of the isle measures 166.4 square miles (431 km2). It is 21 miles (34 kilometers) in length and 14 miles (23 kilometers).'' }\\ \midrule
\incQueryTerms & \textit{``The island of Bimshire is located at 13.4N and 54.4W and is situated in the western area of the North Atlantic Ocean and 100 kilometres east of the Windward Islands and the Caribbean Sea. The island is seen by most scientists as geologically unique as it was formed as a result of an amalgamation of two land masses over a period of many years. The peaceful Arawaks and the more ferocious Caribs were the first inhabitants of Bimshire...''
} \\
\bottomrule
\end{tabular}

\end{table*}

We note that regardless of the ranking function
being biased, publishers are somewhat likely to mimic the documents most highly
ranked for the query in the past \cite{Raifer+al:17a}. With the proposed relevance-model-based biasing,
these documents were highly ranked due to high similarity to the
sub-topic representative documents ($\arbSet$), as discussed
above. Thus, we get a double mimicking/herding effect: making a
document similar to those highly ranked in the past makes the document become more similar to documents
representing the sub-topic of interest.


\section{Empirical Exploration}
\label{sec:eval}

\subsection{Experimental Setting}
\label{sec:expSetup}

To empirically evaluate the content effects presented in
Section~\ref{sec:types} and the techniques proposed in Section
~\ref{sec:means} to drive them, we organized content-based ranking competitions between students in the spirit of those recently used to analyze publishers'
strategies~\cite{Raifer+al:17a}. The competitions were approved by an
international and an institutional ethics committees. The students
who decided to participate signed consent forms and could have opted out at any point.

The competitions are divided to two parts as summarized in
Table~\ref{tab:groups}. Each competition in each part is a repeated
ranking match for a given query that spans five
iterations. A match is described below.
In the first part, we applied the \herding and \biasing approaches to drive the \topical effect; these competitions are referred to as 
\firstmention{\sth} and \firstmention{\stb}, respectively. During
 this part we held additional \firstmention{\control}
 competitions that did not involve any external intervention. In the
 second part, we applied the \herding approach to drive the
 \nonRelevant, \docLength, and \incQueryTerms effects. These
 three types of competitions are denoted \firstmention{\nrh},
 \firstmention{\dlh} and \firstmention{\qth}.

One hundred students in an information retrieval course participated
in the competitions in two different semesters.  Fifty students
participated in each semester. The two-parts structure of the
competitions was identical in both semesters. The quantitative
findings for the two semesters were very similar. We report the
overall quantitative findings over the two semesters.\footnote{Note that each competition in each semester was held separately from the others. The students did not know against whom they were competing as we describe below.} We used $30$ out of the $31$
queries used by Raifer et al.~\cite{Raifer+al:17a}\footnote{Query
  (topic) \#002 was randomly chosen to not be used.}. These queries
were originally selected from the TREC $2009$-$2012$ topic titles
\cite{Raifer+al:17a}; the topics were selected as they had a
commercial intent which was likely to stir up a competition between
the students \cite{Raifer+al:17a}.

In each of the two parts of the competitions, a student was assigned to three queries --- each in a different competition --- which differ between the two parts. In both parts, each student participated in at least two different types of competitions. (There are six types of competitions described in Table~\ref{tab:groups}.)

For each of the \topical-effect types of competitions
(\sth and \stb), two independent competitions were held per query.
Each competition focused on one of the two sub-topics considered for the query which were selected from those of TREC. This resulted in $60$ competitions for the \topical effect.
For all other types of competitions, a
single competition was held per query resulting in $30$ competitions. All in all, we ran 240 competitions of $5$ iterations, each focused on a single query.

Before each competition started, we provided the students with a query
and an initial relevant document.
(Details about the initial documents are provided below.)
In each iteration (match), students were presented with the content and the ranking of the documents submitted in the previous iteration by all the participants in the same competition. The students were incentivized by {\em bonuses} to course grades to modify their documents so as to potentially promote them in the ranking induced in the next iteration. (Students could have received the perfect grade in the course without participating in the competitions.) The documents were plain text of up to $150$ terms.

Two students
participated in each competition.\footnote{Each pair of students competed against each other in at most one competition.}
To maintain lively and dynamic competitions, we artificially added to
each pair of students three additional players so that the students
will see rankings induced over five documents.\footnote{In the herding
  experiments described below, only two players were added as the
  biased document driving the herding was positioned at the first rank
  of the ranking. Hence, there were five documents in each match. Further
  details are provided below.}  Each such ``player'' impersonated one
of the students that participated in
the competition reported in ~\citet{Raifer+al:17a}\footnote{All the documents
  are available at \url{https://github.com/asrcdataset/asrc}.} with
the following exception. If the student remained passive in one of the
iterations and did not modify her document in the competition held in
\cite{Raifer+al:17a}, we randomly selected a document from those
submitted by other students for the same query in the corresponding
iteration.

The students' identities were anonymized
throughout the competitions. Hence, they did not know who their opponents were, and were not aware of the fact that only one of their opponents was a real student.
The analysis presented in Section~\ref{sec:expResults} is based solely on the documents created by the students who actually participated in the competition, and not those created by the additional ``players''.

\paragraph*{Documents}
We used the same initial example document in all the different competitions held for
a query $\query$. We required the initial
document to be relevant --- according to
TREC's topic description ---
to the topic $\topic$ that $\query$ represents and non-relevant to the two selected
sub-topics, $\subTopicParm{1}$ and $\subTopicParm{2}$.
If
a previously published initial document \cite{Raifer+al:17a}
did not meet
these relevance requirements, and could therefore not be used for our competitions, we created a new document using text
snippets that were retrieved for the queries by a commercial
search engine.

For the \herding experiments, we created documents --- to be shown at the
highest rank --- that manifested the four types of content
effects that we wanted to drive as follows.  For the \sth competitions,
we created two biased documents for each query: one of the documents
focused on sub-topic $\subTopicParm{1}$ and the other on sub-topic
$\subTopicParm{2}$.
Hence, we have 60 \sth competitions in total (30 queries $\times$ two competitions per query).
In these competitions we positioned the biased documents at the top of
the presented rankings to drive the \topical effect. The ranking over the
 documents participating in a competition was induced using the LambdaMART ranking function described below. In contrast, in
the corresponding 60 \stb competitions, to drive the same effect, we used a sub-topic
biased relevance model to rank the documents participating in the competition; the relevance model was induced using
Equation~\ref{eq:relModel} for one of the sub-topics using documents relevant to this sub-topic.
We also verified
that the biased document focusing on $\subTopicParm{1}$ was ranked higher
than the biased document focusing on $\subTopicParm{2}$ by a biased relevance
model
induced for sub-topic $\subTopicParm{1}$, and vice versa.  In
addition, in the first iteration before revealing the ranking of
documents to the students, we verified that each biased document would
have been ranked first in the \stb competition when using the
corresponding relevance model.  For the \nrh competitions, we created
non-relevant documents that contained query terms. For \dlh we created
short relevant documents and for \qth relevant documents that did not
contain query terms. Each document that was created for a competition was evaluated by three annotators. Table \ref{tab:docs} shows examples of documents created for the competitions.


\paragraph*{Ranking Functions}
We used Category B of the ClueWeb09 collection with topics $1$-$200$ from TREC $2009$-$2012$ to devise the ranking functions. Topic titles were used as queries. We applied Krovetz stemming to all documents and queries, and removed stopwords on the INQUERY list from queries only. The experiments were performed using the Indri toolkit {(\url{www.lemurproject.org/indri})}.

For all the competitions except for \stb, we followed the
approach used in ~\cite{Raifer+al:17a} to learn a ranking function.
Specifically, we used a learning-to-rank approach, where each
query-document pair was represented by a vector of $26$
content-based features. Most of the features are based on those used
in Microsoft’s learning-to-rank
datasets\footnote{\url{https://tinyurl.com/rmslr}}.  The model was
trained using the top $1000$ documents in a ranking produced using a
language-model-based approach (LM): the negative cross entropy between
the unsmoothed unigram query language model and the Dirichlet-smoothed
(with $\mu=1000$) document language models.  Similarly to Raifer et
al.~\cite{Raifer+al:17a}, we deliberately did not filter out spam
documents, i.e., those assigned a low score by Waterloo's spam
classifier~\cite{Cormack+al:11a}. Instead, we used Waterloo's score as a feature.  Because these scores are not available for the
documents used in our competitions, we did the following.
Five annotators in Figure Eight (\url{www.figure-eight.com}) labeled
each document as valid, keyword stuffed or spam. Then, to simulate
Waterloo's classification scores, we used $20v$, where $v$ is the
number of annotators that marked the document as valid. Since $0 \le v
\le 5$, the scores we get are in $[0,100]$, as is the case for the
original Waterloo scores.  Thus, the ranking model was trained using
Waterloo's spam scores, and applied using the human-created scores.

We used LambdaMART~\cite{Wu+al:10a} via the RankLib library ({\url{https://sourceforge.net/p/lemur/wiki/RankLib}}) to learn the model. We randomly split the queries into four folds. Three folds were used to train the model and the remaining fold to set hyper-parameter values; NDCG@5 served as the optimization metric. The number of trees and leaves in LambdaMART were set to $250$ and $50$, respectively, following experiments with values in $\set{250,500}$ and $\set{5,10,25,50}$.

The relevance model (\RM) for the \stb competitions (Equation
~\ref{eq:relModel}) was constructed from five randomly sampled
  relevant documents per sub-topic. To set the number of expansion
  terms, we created relevance judgments as follows. The five documents
  from which the relevance model was constructed were considered
  ``relevant''. Five documents that were relevant to the topic, but
  not to the sub-topic in question and were assigned the highest
  LM score (see above) were considered ``non-relevant''. The number of
  expansion terms was set to $100$ to optimize the NDCG@5 of a ranking over the
  ten ``judged'' documents, following experiments with values in
  $\set{10,25,50,100}$.

All the documents created by students throughout the competitions were judged for relevance with respect to a query's topic by five annotators in Figure Eight. In addition, documents created in the \topical-effect competitions were also judged for relevance with respect to the two selected sub-topics.
For the STB and STH competitions, queries \#144 and \#164 were removed from the final analysis due to lack of available sub-topics.

The statistical significance of the difference between two {\em sets}
of $30$ competitions (each held for a different query) with respect to a
measure/effect is determined using a paired permutation
(randomization) test with $p=0.05$; $100000$ permutations were
randomly sampled. Pairing is done with respect to queries and
iterations. The value of the measure/effect per query and iteration is
the average value for the two documents of the two participating
students. Bonferroni correction was applied for multiple comparisons.

\subsection{Experimental Results}
\label{sec:expResults}

\subsubsection{The \topical Effect}
\label{sec:topicResults}

To drive the \topical effect we used the \herding and \biasing
approaches in the \sth and \stb competitions, respectively. For a given query, we
selected two sub-topics and held a competition with respect to
each. Only one sub-topic was ``active'' in a given competition; i.e., the
sub-topic was the focus of the highest ranked document in \sth (which was biased) or
of the biased relevance model (Equation~\ref{eq:relModel}) used for document ranking in \stb. The second
selected sub-topic in this competition was ``passive''; i.e., no ``driving'' with respect to this sub-topic was performed. 
\firstmention{\act} and \firstmention{\pass} denote the active and
passive sub-topics. Thus, for each topic, there was one competition where one sub-topic was active and the other one passive and one competition in which the reverse holds. Accordingly, we have equal representation for each of the two sub-topics of a topic as active and passive in the competitions.

To measure the extent of documents becoming
focused on a sub-topic, we can measure the relative
similarity of their induced language models to the sub-topic biased
relevance models.\footnote{In both \sth and
  \stb, the documents became more similar to the active sub-topic than
  to the passive sub-topic in terms of language models. These results are
  omitted as they exhibited similar patterns, although to a somewhat less emphasized extent, than those presented below.} However, the sub-topic biased relevance models also encode information about the topic as a whole. To
 distill the information specific to the sub-topic with respect to the entire topic, and
use that information to measure similarities to sub-topics, we utilized a two component mixture model described in Appendix \ref{sec:app} to induce a {\em distilled sub-topic model}.
Then, similarity of a document to a sub-topic is measured based on the negative cross entropy between the distilled sub-topic model and the document language model.
In addition, we analyze the
cosine similarity between the TF-IDF vectors representing a student's document and the biased (planted) documents in the \sth competitions.
The different similarity scores are averaged over
documents per match and over queries per iteration.
We present for reference the results for the \control
competitions where the similarity scores to both sub-topics were averaged.
\begin{figure}[t]
\centering
\begin{adjustbox}{max width=\textwidth}
\begin{tabular}{cc}
    \includegraphics[height=\figHeight,width=\figWidth]{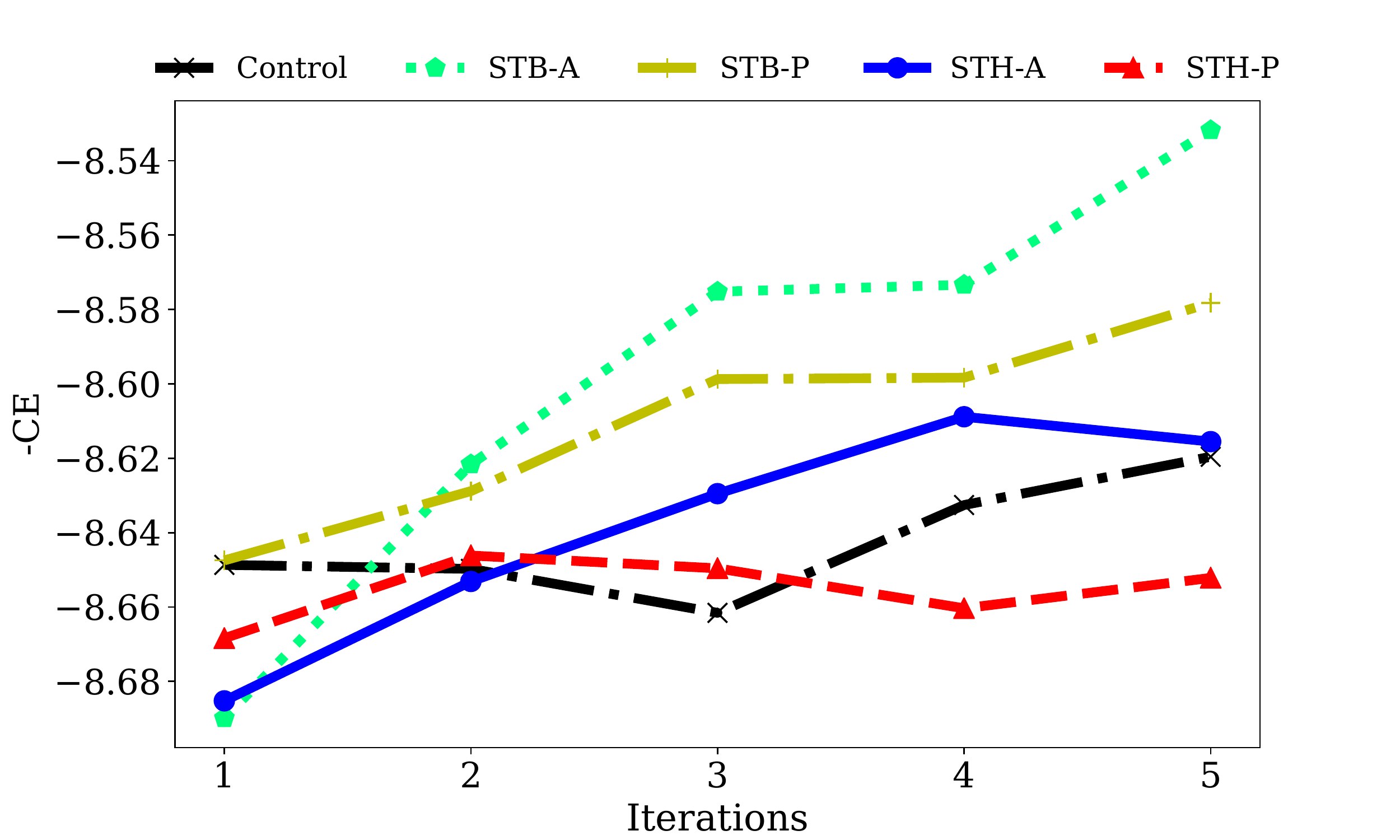} \\
    \includegraphics[height=\figHeight,width=\figWidth]{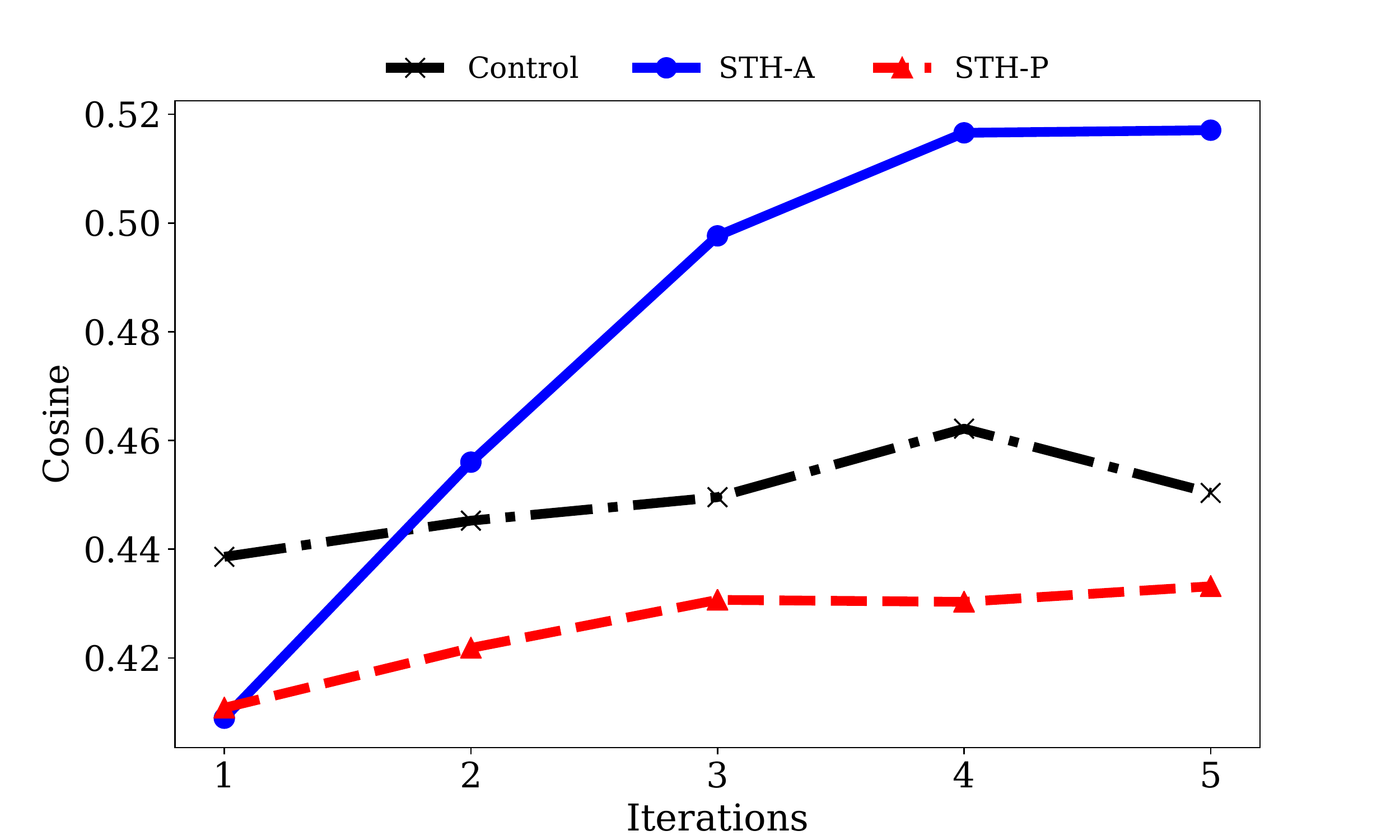}
\end{tabular}
\end{adjustbox}
\caption{\label{fig:topic} The \topical effect. Top figure: the average language model similarity (negative cross entropy) between a document and a distilled sub-topic model in the STH and STB competitions. \{STH,STB\}-\{A,P\} refers to the similarity in the STH/STB competitions to the distilled model induced for the active (A) or passive (P) sub-topic. Bottom figure: average cosine similarity with the documents biased for the active (STH-A) and passive (STH-P) sub-topics in the STH competitions. Both figures: in the Control competitions, the average similarity is computed for both sub-topics (A and P). In terms of -CE,  STB-A is statistically significantly different from STB-P and Control. In terms of Cosine, STH-A is statistically significantly different from Control and STH-P.}
\end{figure}
\begin{figure}[t]
\centering
\begin{adjustbox}{max width=\textwidth}
\begin{tabular}{cc}
\includegraphics[height=\figHeight,width=\figWidth]{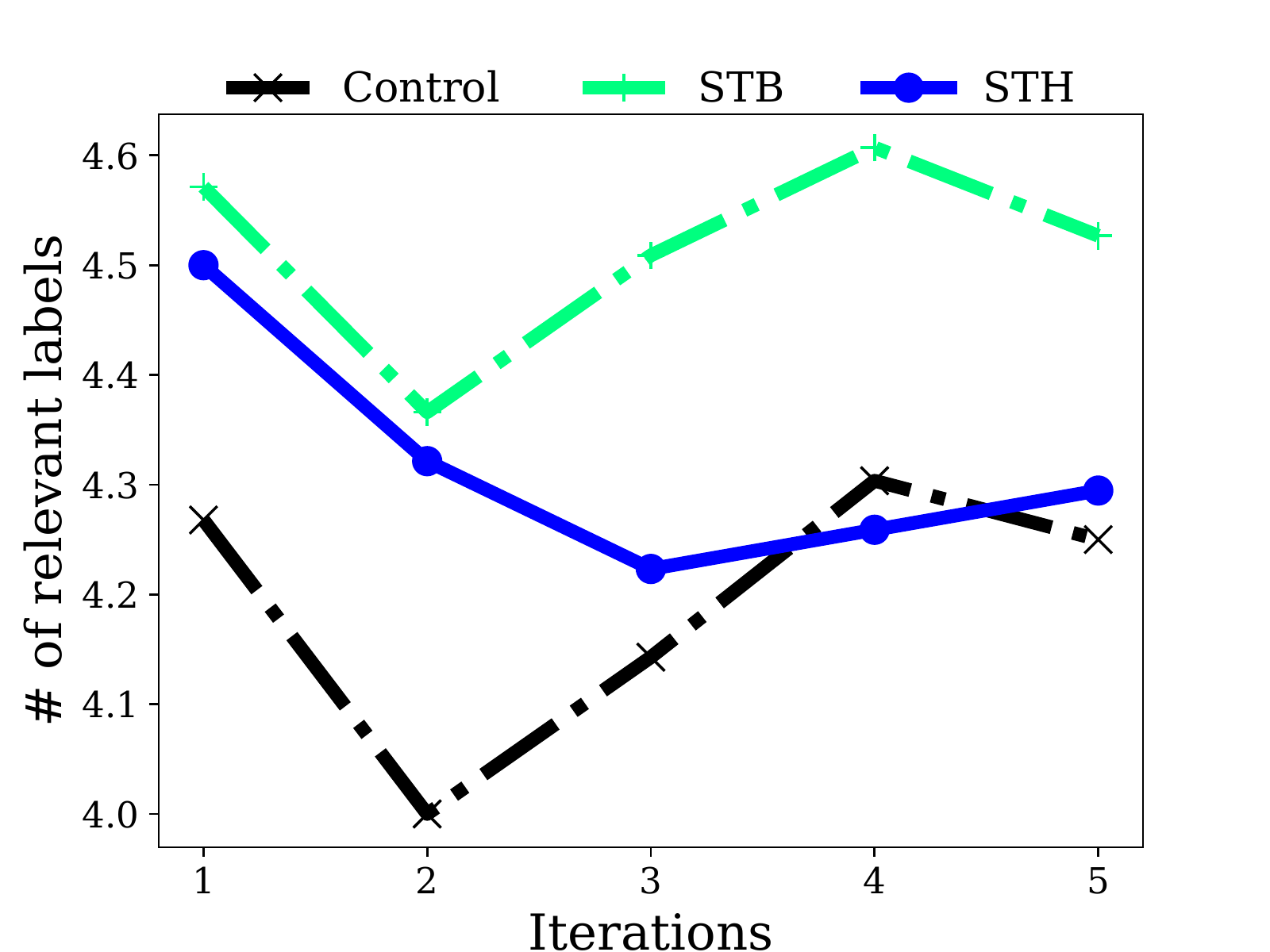} \\
\includegraphics[height=\figHeight,width=\figWidth]{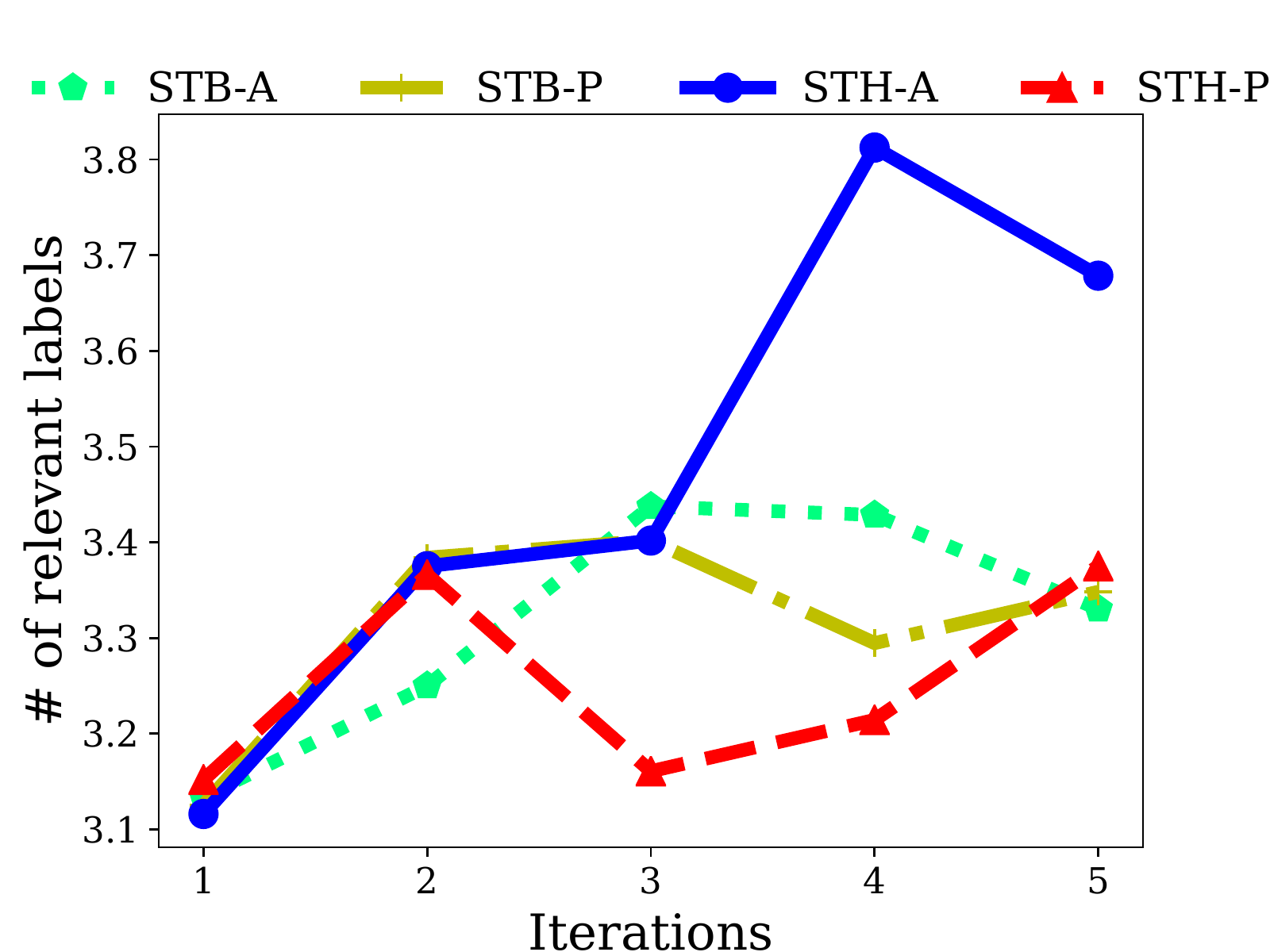}
\end{tabular}
\end{adjustbox}
\caption{\label{fig:topicRel} The average number of relevant labels assigned to documents with respect to a query's topic (top) and the two sub-topics (bottom). STB is statistically significantly different from both Control and STH. STH-A is statistically significantly different from STH-P and STB-P.}
\end{figure}

We see in Figure \ref{fig:topic} that most
similarities increase along the iterations regardless of the
distilled models or biased documents used. Even in the \control
competitions, which did not involve any external intervention, a
general moderate upward trend is observed. The average language-model-based
similarity of a document to a distilled active sub-topic is almost
always higher than that for the distilled passive sub-topic (\stba
vs. \stbp and \stha vs. \sthp in the top figure); the gap between
these two similarities almost always increases along the iterations
and is, on average, statistically significant for the STB competitions. The language-model-based similarity for \stbp
is also higher (in a statistically significant manner) than the average similarity to both sub-topics in the
Control group.\footnote{In the Control groups there is no herding or biasing of the ranking function. Hence, similarities are computed for both sub-topics and are then averaged.} These findings suggest that the topics of the documents
created by students gradually shifted towards the active sub-topic to
a larger extent compared to the passive sub-topic in both approaches for driving the \topical effect (\herding and \biasing). In comparing \stba with \stha, we see that the former posts higher similarities which means that the \biasing approach is more aggressive in driving the \topical content effect than the \herding approach.

We also see in Figure~\ref{fig:topic} (bottom) that in the \sth
competitions, documents written by students are much more similar to
the biased document which focuses on the active sub-topic and which is
shown at the top of the ranking than they are to the document biased
to the passive sub-topic which they were not shown (\sth-\act
vs. \sth-\pass). The differences are statistically significant. These
findings further attest to a herding effect.


To summarize, both the \herding and \biasing approaches were effective in driving
the \topical effect with the latter being somewhat more effective.


\myparagraph{Relevance} As noted in Section~\ref{sec:expSetup}, all
the documents created by students for a query were annotated by five
annotators for relevance with respect to the query's topic and the two
sub-topics. Figure~\ref{fig:topicRel} presents the average number of
relevant labels (per topic and sub-topic) assigned to a document.

We see in Figure \ref{fig:topicRel} (top) that in the \sth
competitions, the relevance of documents to the query's
topic (top figure) substantially decreased until iteration three and
then rised a bit, but to a level quite lower than that at the
first iteration. In contrast, in the \stb competitions, the relevance
level at iteration five was almost as that as at the first iteration, albeit
fluctuations along the iterations.


Figure \ref{fig:topicRel} (bottom) shows that the relevance to both
sub-topics, in both STH and STB, overall
increased from iteration one to iteration five. This is in line with
the sub-topic-based similarity findings presented above. The most
prominent increase is for STH-A whose results are statistically significantly higher than those for STH-P. In contrast, the difference between STB-A and STB-P is not statistically significant. Thus, while both the \herding and \biasing approaches drove the \topical effect, the former also helped more in increasing relevance for the target (active) sub-topic.



\begin{figure}[t]
\centering
\includegraphics[height=\figHeight,width=\figWidth]{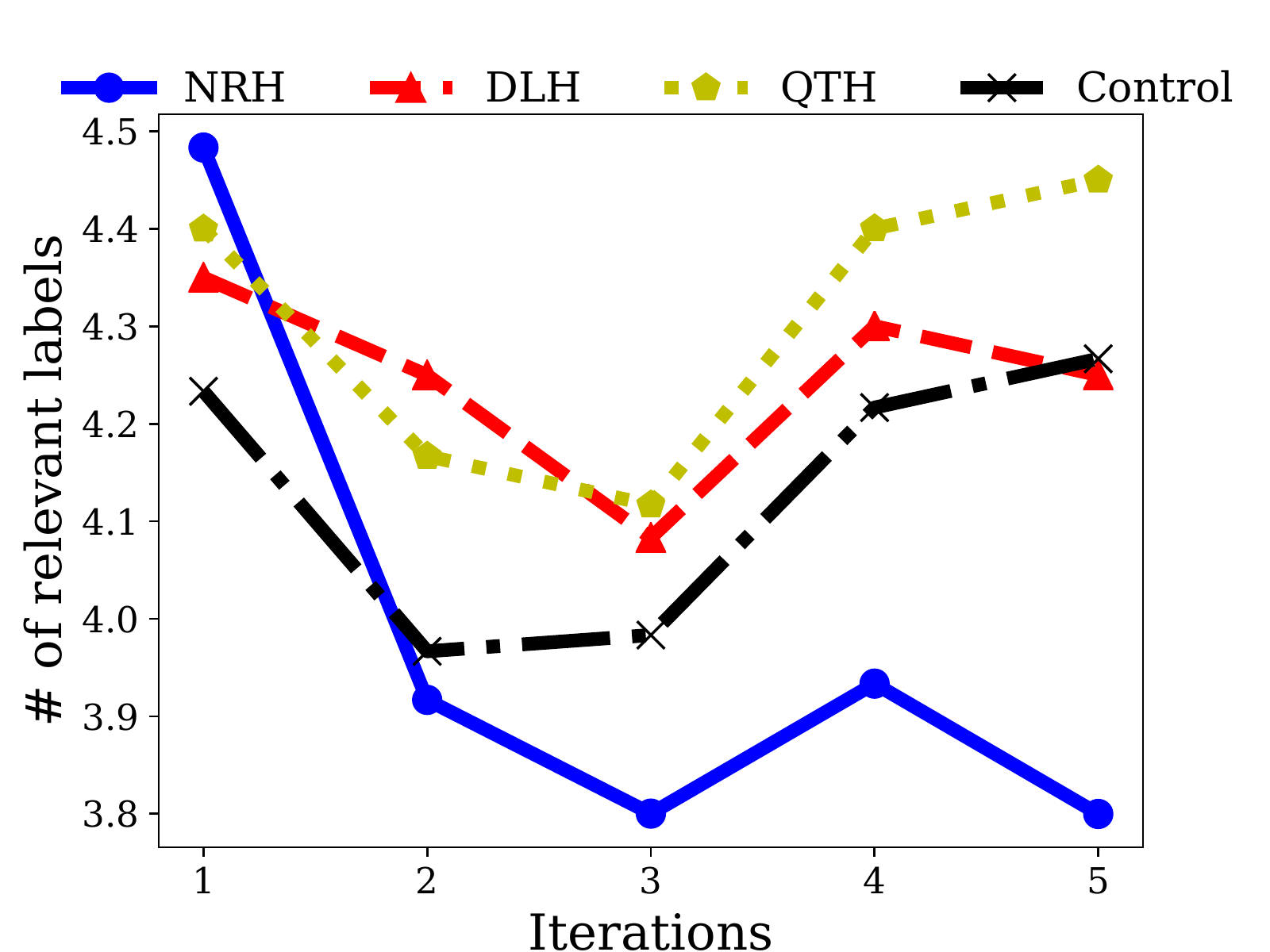}
\caption{The \nonRelevant effect: average number of relevant labels per document and iteration. NRH is statistically significantly different from DLH, QTH and Control.}
\label{fig:nonrel}
\end{figure}

\subsubsection{The \nonRelevant Effect}

The \nonRelevant effect was driven in the \nrh competitions using the
\herding approach. Non-relevant documents were positioned at the top
of the presented rankings.  Figure~\ref{fig:nonrel} presents the
average number of relevant labels assigned to the students' documents
in each iteration. For reference, we show the results for the (i)
\control competitions, in which no external interventions were
performed, and (ii) the \dlh and \qth competitions, in which two other
content effects were driven using the \herding approach. In the
\control, \dlh and \qth competitions we observe fluctuations in the
average number of relevant labels assigned to documents; but, the
number for the first iteration is not very different than that for the
last iteration. Recall that the students had no incentive to produce relevant documents, but rather documents that are highly ranked. 

Strikingly, we see in Figure \ref{fig:nonrel} that for \nrh there is a sharp 
decline in the number of relevant labels which results in a statistically significant difference with all other three competitions. This finding shows that using a rather simple \herding
approach, a search engine can lead to a substantial reduction of the amount of
relevant content for a specific topic in the corpus.

\begin{figure}[t]
\centering
\includegraphics[height=\figHeight,width=\figWidth]{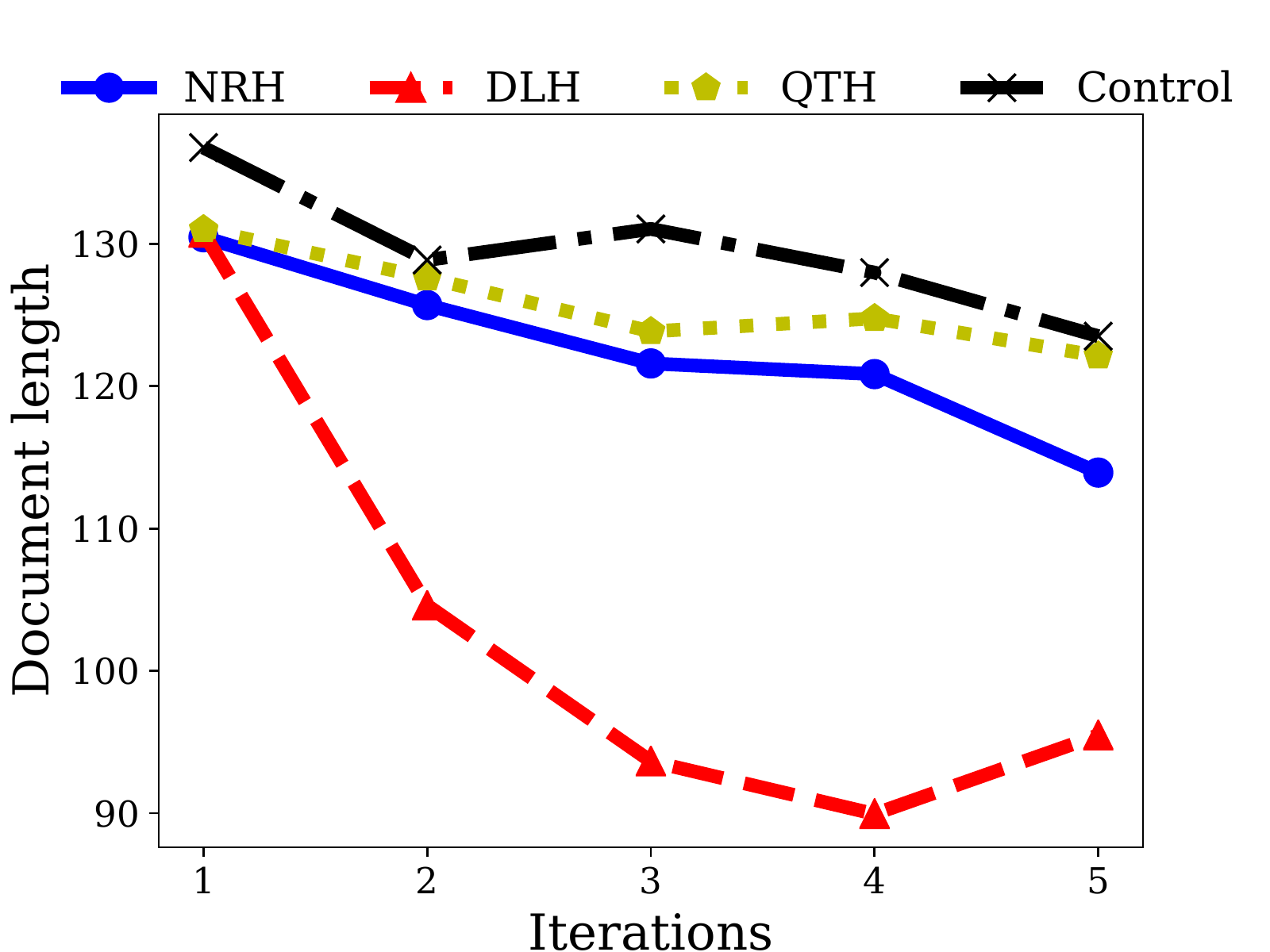}
\caption{The \docLength effect. DLH is statistically significantly
  different from NRH, QTH and Control.}
\label{fig:len}
\end{figure}

\begin{figure}[t]
\centering
\begin{adjustbox}{max width=\textwidth}
\begin{tabular}{c}
\includegraphics[height=\figHeight,width=\figWidth]{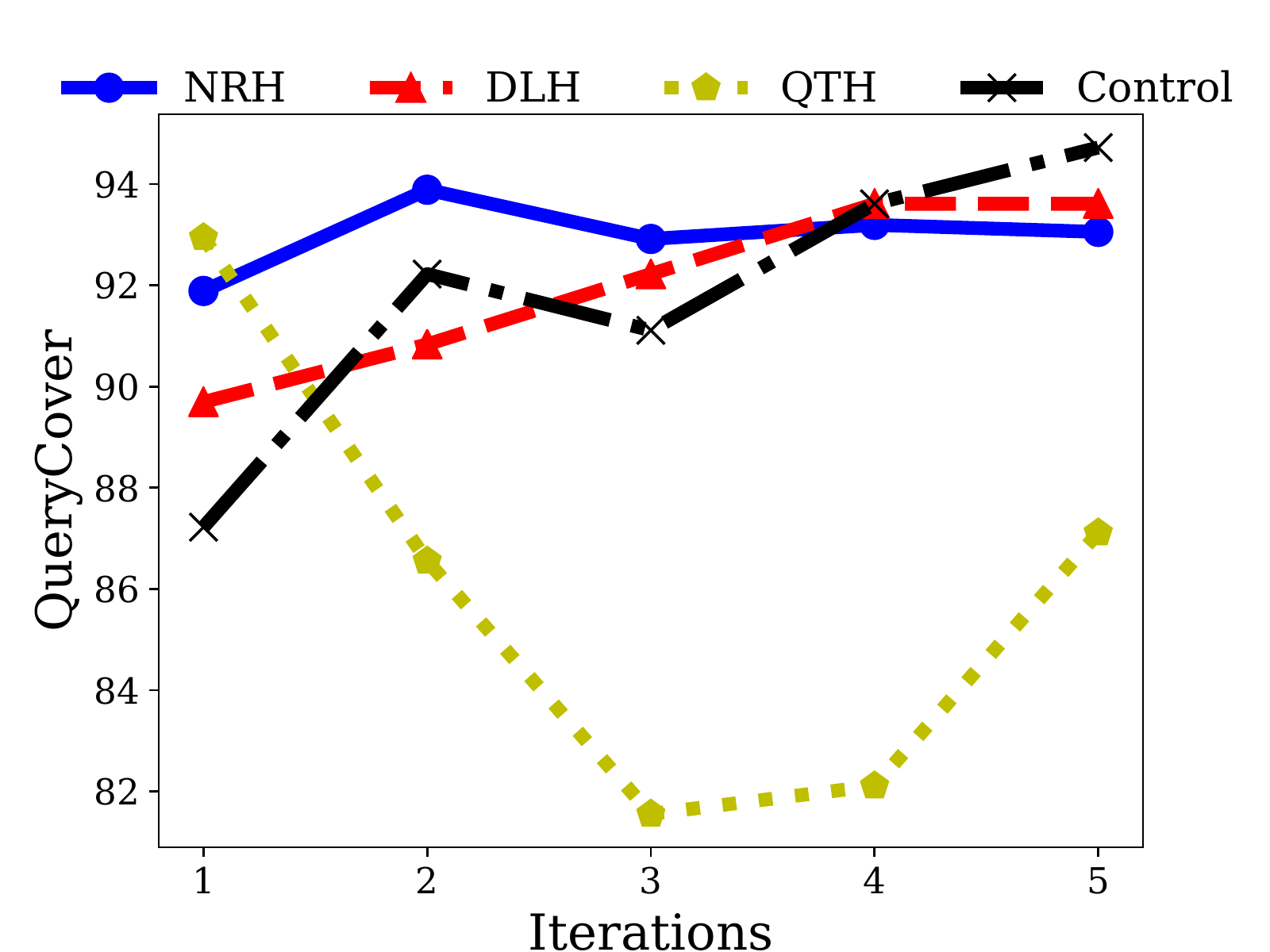} \\
\includegraphics[height=\figHeight,width=\figWidth]{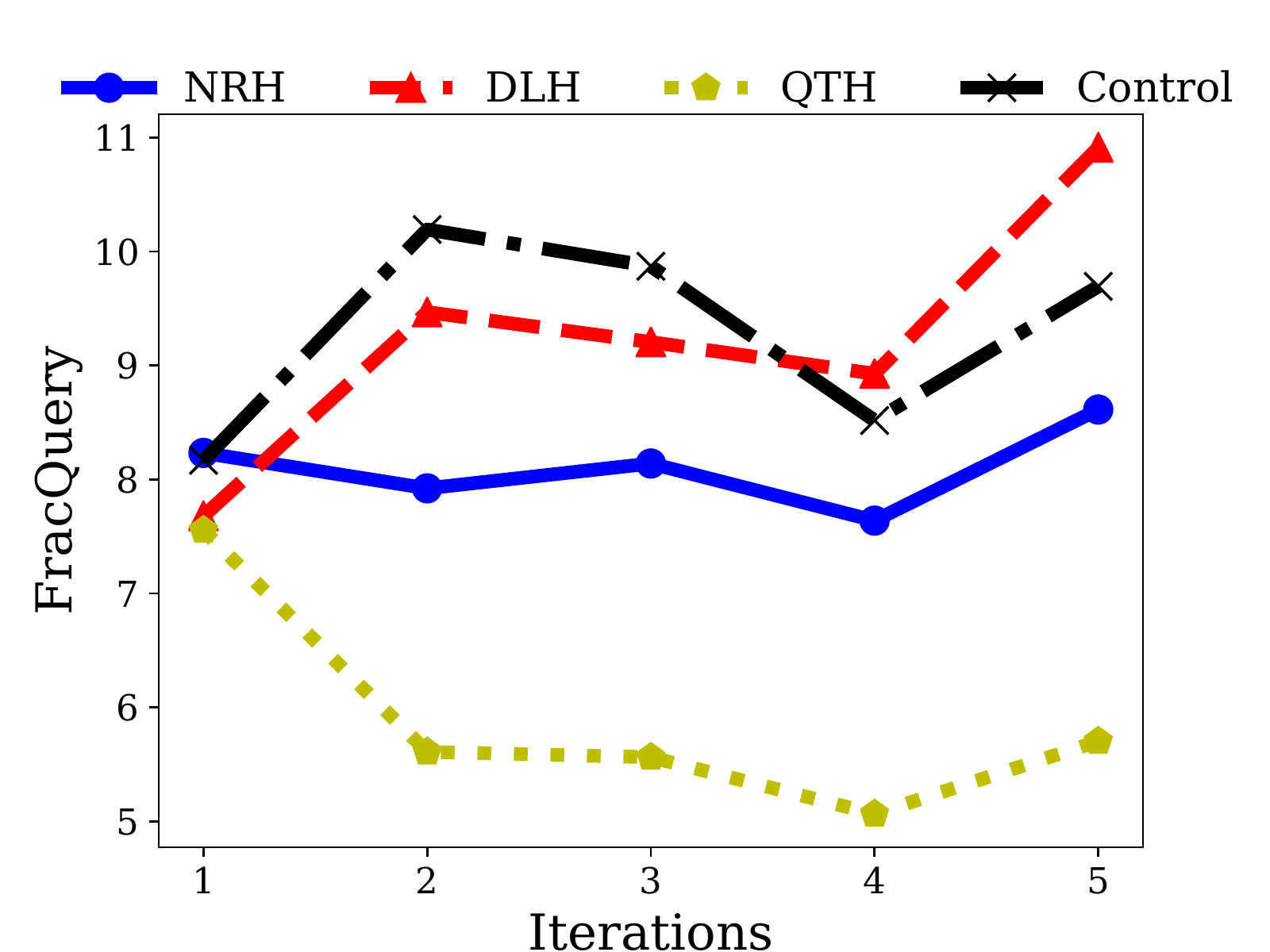}
\end{tabular}
\end{adjustbox}
\caption{\label{fig:query} The \incQueryTerms effect. We present for each iteration the average percentage of query terms appearing in a document (\queryCover) and the average percentage of terms in a document that are query terms (\fracQuery). QTH is statistically significantly different from NRG, DLH and Control for both \queryCover and \fracQuery.}
\end{figure}

\subsubsection{The \docLength Effect}

Thus far, we showed that search engines can affect the coverage of
topics in a corpus. We now turn to study the ability of search engines
to drive surface-level changes in the corpus. We examine the \dlh
competitions in which short relevant documents were positioned at the
top of the rankings to drive the \docLength effect.

Figure~\ref{fig:len} shows for each
iteration the average length of students' documents across the queries. We see that in the \dlh competitions, the document
length sharply decreased during the first four iterations. On average,
the document length in these competitions was statistically
significantly lower than that in the other three competitions which
attests to a clear herding effect. The mild increase of document
lengths in the fifth iteration of the \dlh competitions can
potentially be explained as follows: students started noticing
that further shortening the documents does not help them to promote
them for two reasons: (i) the top most document throughout the iterations was the
biased one we planted, and (ii) the ranking function we employed does not necessarily reward short documents.


We also see in Figure \ref{fig:len} a gradual decrease of document
length in the \qth, \control and \nrh competitions, with the latter
exhibiting the largest decrease of the three. The finding about NRH can be
explained by the fact that the non-relevant documents we planted to
drive the \nonRelevant effect via herding, were on average shorter
than the initial relevant documents provided to the students as examples ($118.6$ terms vs. $133.1$ terms). Hence, there was a double herding effect in the \nrh competitions: documents were made shorter and less relevant. Still, the length decrease is statistically significantly smaller than that for DLH where the planted documents were of an average length $30.3$ terms. For \qth and \control the planted documents were of similar length to that of the example relevant documents, which could potentially help to explain the very mild length decrease in these competitions.

To summarize, the shorter the documents we posted at the top of the
ranking, the shorter the documents
created by the students were. This finding attests to a clear herding effect.

\subsubsection{The \incQueryTerms Effect}

To drive the \incQueryTerms effect in the \qth competitions, relevant documents that do not contain query terms were positioned first in the rankings presented to students. In Figure~\ref{fig:query} we show the percentage of query terms that appear in a document (\queryCover) and the percentage of terms in a document that are query terms (\fracQuery). The values are averages over documents and queries per iteration.

Figure \ref{fig:query} shows that in the \qth competitions, there is a
substantial downward trend for both \queryCover and \fracQuery along
the first few iterations and then some increase in the last iteration
or two. The initial decrease together with the --- statistically
significantly different --- upward trend observed for NRH, DLH and
Control, attests to an herding effect in the QTH competitions. This
finding is quite striking: while the participating students knew about
the importance of query-terms occurrence in documents as a relevance
signal in virtually all retrieval methods, they inferred that the biased
document was ranked first due to not having query terms; hence, they reduced query term occurrences in their documents.

The increase of values in the last iteration or two in the QTH competitions can be explained as
follows. The ranking function used in all the herding-based competitions rewards query
terms occurrence as the features it uses are based on textual
document-query similarities.  In fact, \queryCover is one of the
features used in the learning-to-tank model. (Refer back to
Section~\ref{sec:expSetup} for details.) This explains, for example,
the upward trend for the NRH, DLH and Control competitions. Now,
presumably, students that removed query terms from their documents in
the first few iterations of the QTH competitions following the planted document, did not manage to
promote their documents in rankings. On the contrary, some of these documents were demoted.


\section{Discussion}
\label{sec:discuss}
Heretofore, we have focused on the ability of search
engines to drive pre-defined, targeted, content effects. There is obviously no reason for search engines to employ such practices. However, this ability can potentially be abused by publishers as we discuss next. 

Say that a
publisher is interested in promoting a content effect in the corpus. She can then write a document
that manifests the effect and try to optimize it with respect to the
ranking function. If successful in having her document ranked first, the publisher can
potentially affect the content other publishers produce due to the herding phenomenon.

In the experiments we reported in Section
\ref{sec:eval}, documents we planted at the first rank which manifested specific content effects were manually selected
and modified to this end. However, we argue that with the progress in language
generation capabilities based on pre-trained language models (e.g.,
BERT \cite{Delvin+al:18a}, XLNet
\cite{Yang+al:19a}, GPT3 \cite{brown2020language}, etc.), this challenge will become easier along
time. For example, ``high quality'' fake news were generated using
advanced language generation techniques \cite{Zellers+al:19a}.


\section{Conclusions and Future Work}

We presented a first study of simple techniques that search engines
can employ to drive pre-defined targeted content effects in the corpus via the rankings they
induce. The first is based on the {\em herding} phenomenon from the
economics literature, and the second is based on biasing the ranking
function. We explored topical effects and several document-property effects. Analysis of content-based ranking competitions we organized demonstrated the ability of a search engine to drive these effects using the suggested techniques. The concern is not that search engines will actually deploy such techniques, but rather that documents' authors will use the engines as platforms to applying such techniques.

For future work we plan to study herding effects, both theoretically
and empirically, when publishers optimize their documents for
multiple queries; i.e., going beyond the single-query setting addressed here and by Raifer et al. \cite{Raifer+al:17a}.

\myparagraph{Acknowledgments}
We thank the reviewers for their comments.
The work by Moshe Tennenholtz and Gregory Goren was supported by funding from the European Research Council (ERC) under the European Union’s Horizon 2020 research
and innovation programme (grant agreement 740435).

\appendix
\section{Distilling sub-topic models}
\label{sec:app}
To distill a unigram language model, $\theta_{\subTopicParm{s}}$, that represents the unique information in a sub-topic $\subTopicParm{s}$ with respect to the more general topic $\topic$, we use a mixture model (cf., \cite{Zhai+Lafferty:01a}):
$$\log \mathcal{L} \definedas \sum_{\doc_i \in \subTopicParm{s}}
\sum_{w \in \doc_i} tf (w;d_i) \log ((1-\lambda)
\condP{w}{\theta_{\subTopicParm{s}}} + \lambda
\condP{w}{\topic}) ;$$ ${\mathcal L}$ is the likelihood
function; $\doc_i \in \subTopicParm{s}$ are all documents (in TREC's qrels files) marked as
relevant to $\subTopicParm{s}$; $w$ is a term; $tf(w;d_i)$ is the number of times $w$ appears in $d_i$; $\lambda$ is a free parameter; $\condP{w}{\topic}$ is the probability assigned to term $w$ by a maximum likelihood estimate induced from all documents marked as relevant to $\topic$ (in TREC's qrels files). We use the EM algorithm to infer
$\theta_{\subTopicParm{s}}$ --- i.e., set
$\condP{w}{\theta_{\subTopicParm{s}}}$ for each term $w$. 

To compute the cross-entropy-based similarities between a distilled
sub-topic model $\theta_{\subTopicParm{s}}$ and a document language
model (see Section \ref{sec:topicResults}), we clip (and normalize)
$\theta_{\subTopicParm{s}}$ (cf. \cite{Zhai+Lafferty:01a}) to use the
$\alpha$ terms to which it assigns the highest probability. 
The procedure that was used to set the number of expansion terms in the relevance model (see Section~\ref{sec:expSetup})
was also used to set
$\alpha$ ($\in \set{10,25,50,100}$) and $\lambda$ ($\in
\set{10,25,50,100}$).




\balance
\clearpage
\newpage
\bibliographystyle{ACM-Reference-Format}
\bibliography{cunlp-ir}
\end{document}